# Lattice Vibrational Modes and Phonon Thermal Conductivity of Monolayer MoS$_2$

Yongqing Cai, Jinghua Lan, Gang Zhang* and Yong-Wei Zhang

Institute of High Performance Computing, A*STAR, Singapore 138632
*Email: zhangg@ihpc.a-star.edu.sg

**ABSTRACT:** The anharmonic behavior of phonons and intrinsic thermal conductivity associated with the Umklapp scattering in monolayer MoS$_2$ sheet are investigated via first-principles calculations within the framework of density functional perturbation theory. In contrast to the negative Grüneissen parameter ($\gamma$) occurring in low frequency modes in graphene, positive $\gamma$ in the whole Brillouin zone is demonstrated in monolayer MoS$_2$ with much larger $\gamma$ for acoustic modes than that for the optical modes, suggesting that monolayer MoS$_2$ sheet possesses a positive coefficient of thermal-expansion. The calculated phonon lifetimes of the infrared active modes are 5.50 and 5.72 ps for E′ and A$_2''$ respectively, in good agreement with experimental result obtained by fitting the dielectric oscillators with the infrared reflectivity spectrum. The lifetime of Raman A$_1'$ mode (38.36 ps) is about 7 times longer than those of the infrared modes. The dominated phonon mean free path of monolayer MoS$_2$ is less than 20 nm, about 30-fold smaller than that of graphene. Combined with the non-equilibrium Green's function calculations, the room temperature thermal conductivity of monolayer MoS$_2$ is found to be around 23.2 Wm$^{-1}$K$^{-1}$, two orders of magnitude lower than that of graphene.

## I. INTRODUCTION

As a semiconducting analogue of graphene, the atomically thin MoS$_2$ monolayer, consisting of a hexagonal lattice of Mo atoms sandwiched between two similar lattices of S atoms in a trigonal prismatic arrangement, has recently attracted considerable attention for field effect transistor (FET) and optical device applications due to its presence of a finite bandgap [1, 2]. Great efforts have been made to understand the dynamics of carriers of MoS$_2$ including mobilities of excitons [3, 4], electrons/holes [5, 6] and the effects from electrical [7] and stress [8-10] fields applied within the layers. In contrast to graphene, dielectric screening associated with the electron-electron interaction and electron-phonon coupling of the inter- and intra-layers gives rise to an anomalous frequency shift [11], a symmetry-dependent phonon renormalization [12], and a superconducting behavior in MoS$_2$ [13].

In addition to the potential applications in FET devices, MoS$_2$ has recently shown an intriguing capability of a thermoelectric energy conversion, where a large value of the Seebeck coefficient for single-layer MoS$_2$ ($-4 \times 10^2$ and $-1 \times 10^5$ $\mu$VK$^{-1}$ depending on the strength of the external electric field) was demonstrated [14]. For both FET and thermoelectric applications, the phonon property of monolayer or few-layer (FL) MoS$_2$ is critical. On the one hand, the electron-acoustic phonon coupling dominates the scattering of low-energy carriers, which limits the carrier mobility [5,6]. On the other hand, monolayer MoS$_2$ is a semiconductor with a sizable bandgap. Thus the electrons have limited contribution to thermal conductivity ($\kappa$), and the intrinsic $\kappa$ is dominated by phonon contribution. Recent molecular dynamics (MD) simulations showed that the $\kappa$ value for monolayer MoS$_2$ is 1.35 Wm$^{-1}$K$^{-1}$ by Liu *et al.* [15] and 6 Wm$^{-1}$K$^{-1}$ by Jiang *et al.* [16] Experimental measurements of the $\kappa$ of FL MoS$_2$ were reported to be between 0.4-1.59 Wm$^{-1}$K$^{-1}$ [17,18]. More recently, using Raman spectroscopy approach, Sahoo *et al.* reported a value of around 52 Wm$^{-1}$K$^{-1}$ for FL MoS$_2$ [19]. Compared to the extensive studies on the thermal conductivity of grapheme [20-23], comprehensive analysis of the Grüneissen parameter, phonon relaxation time and phonon mean free path (MFP) of MoS$_2$ is currently still lacking, despite their critical role in the understanding of the phonon scattering, temperature effect, and phonon-mode contribution to the intrinsic thermal conductivity.

In this study, by using density functional perturbation theory (DFPT), we investigate the lifetime of phonons, and intrinsic $\kappa$ of monolayer MoS$_2$ by calculating the Grüneissen parameters, frequency- and polarization-dependent phonon relaxation time and MFP. Positive Grüneissen parameters for all the modes, in contrast to the negative Grüneissen of the low-frequency mode in graphene, are found for monolayer MoS$_2$. The calculated dominated MFP of MoS$_2$ is around 18.1 nm. Based on non-equilibrium Green's function (NEGF) scheme, the intrinsic $\kappa$ at room temperature is found to be around 23.2 Wm$^{-1}$K$^{-1}$. Our study shows that, owning to the S–Mo–S sandwich structure, the single-layer MoS$_2$ is dramatically different from the one-atom-thick graphene with respect to the structural stability, thermal expansion, vibrating anharmonic behavior and thermal conductivity.

## II. COMPUTATIONAL METHOD

The calculations of the interatomic force constants (IFC) and phonon dispersion are performed using the Quantum-Espresso code [24], within the local density approximation (LDA) of Perdew-Wang. We use the norm-conserving pseudopotential with energy (charge density) cutoff up to 70 Ry (700 Ry). The first Brillouin zone is sampled with a $30 \times 30 \times 1$ Monkhorst-Pack grid. The vacuum region thickness is greater than 16 Å. The structures are relaxed until the forces exerted on the atoms are less than 0.01 eV/Å. The optimized equilibrium lattice constant of monolayer MoS$_2$ is 3.09 Å, smaller than the measured value of 3.16 Å [25] as LDA normally underestimates the lattice constant. In the following DFPT calculation, a Monkhorst-Pack $10 \times 10 \times$ 1 q-mesh is used to calculate the dynamical matrix at each q grid, which gives the IFC through inverse Fourier transform to real space.

The thermal conductance is calculated based on NEGF approach [26,27]. The ballistic thermal conductance of a junction connected to two leads at different equilibrium heat-bath temperatures is given by the Landauer formula,

$$\sigma(T) = \int_0^\infty \hbar\omega T[\omega] \frac{\partial f_B(\omega,T)}{\partial T} \frac{d\omega}{2\pi} \quad (1)$$

where $f_B(\omega,T) = 1/(e^{\hbar\omega/k_B T} - 1)$ is the Bose-Einstein distribution function for a phonon with a frequency at the reservoirs, $T[\omega]$ is the transmission coefficient, $\hbar$ is Planck's constant, and $T$ is the average temperature of the hot and cold baths. Within the framework of NEGF, the phonon-transmission function $T[\omega]$ is given by $T[\omega] = Tr[G^r \Gamma_L G^a \Gamma_R]$, where $G^r$ and $G^a$ are respectively the retarded and advanced Green's functions of the central region connected with two leads defined as $G^r = (G^a)^+ = [\omega^2 - K^c - \Sigma_L^r - \Sigma_R^r]^{-1}$ with $K^c$ being the force constant matrix and L(R) denoting the left (right) leads, $\Gamma_\alpha$ ($\alpha$ = L and R) is the broadening function describing the ability of phonons to enter and leave the leads and given by $\Gamma_\alpha = i(\Sigma_\alpha^r - \Sigma_\alpha^a)$, where $\Sigma_\alpha^r$ and $\Sigma_\alpha^a$ are the self-energies of the leads accounting for coupling of the central part with the leads. Here the IFC is directly calculated by first-principles.

### III. RESULTS AND DISCUSSION

#### A. Phonon dispersion

The lattice dynamics of bulk and monolayer $MoS_2$ have been studied both experimentally [11, 28-30] and theoretically [31,32]. To facilitate the discussion of the mode-dependent scattering behavior and comparison with experimental results, as inspired by the previous work [33], we summarize the characters of the phonons for both the bulk (2H phase) and monolayer $MoS_2$ in Figure 1a with respect to the symmetry assignment, frequency, optical character, and eigenvectors. Since the primitive cell of $2H-MoS_2$ and monolayer $MoS_2$ contains 6 and 3 atoms, there are a total of 18 and 9 phonon modes, respectively. A factor group analysis of the point group ($D_{6h}$ and $D_{3h}$ for $2H-MoS_2$ and monolayer $MoS_2$, respectively) shows that the long-wavelength optical phonon modes at the $\Gamma$ point (without the three translational acoustic modes) can be decomposed as

$\Gamma_{Optical}(2H - MoS_2) = A_{2u}(IR) + E_{1u}(IR) + A_{1g}(R) + 2E_{2g}(R) + E_{1g}(R) + 2B_{2g}(IN) + B_{1u}(IN) + E_{2u}(IN);$ (2a)

$\Gamma_{Optical}(monolayer) = A_2''(IR) + E'(IR + R) + A_1'(R) + E''(R)$ (2b)

where all the Raman (R), infrared (IR), and inactive (IN) modes are assigned. Note that the $A_{2u}$, $A_{1g}$, $B_{2g}$, $B_{1u}$, $A_2''$, $A_1'$ modes are singly degenerate and the $E_{1u}$, $E_{2g}$, $E_{1g}$, $E_{2u}$, $E'$, $E''$ modes are doubly degenerate. The R and IR modes are mutually exclusive in $2H-MoS_2$ due to the presence of inversion symmetry in the crystal. The two IR modes in $2H-MoS_2$, $A_{2u}$ and $E_{1u}$, evolve into the IR active $A_2''$ and $E'$ modes in the monolayer case, respectively, where the latter is also Raman active due to the lack of inversion center in monolayer and assigned as $E_{2g}^1$ in bulk or FL $MoS_2$ [11]. Another out-of-plane $A_{1g}$ Raman mode in $2H-MoS_2$, which is normally used to identify the layer number of FL $MoS_2$ flakes, matches with the lar $A_1'$ mode in the monolayer case, where the top and bottom sulfur layers vibrate out-of-phase with direction normal to the basal plane while the Mo layer remains stationary. Our calculated frequencies for these modes listed in Fig.1(a) are in good agreement with experimental measurements [1,2].

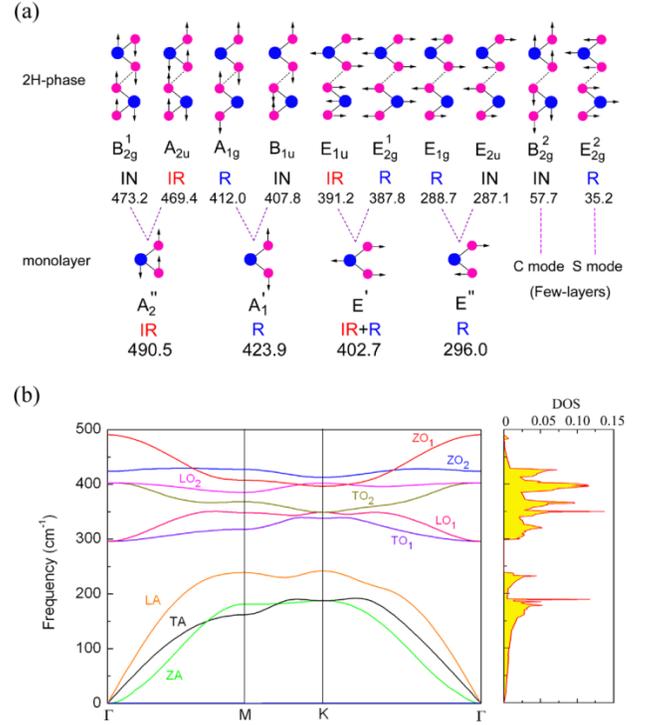

FIG. 1. (a) Comparison of the zone-center modes between $2H-MoS_2$ and monolayer $MoS_2$. The frequencies of the 2H phase are adopted from Ref. [32]. (b) Phonon dispersion and phonon DOS for monolayer $MoS_2$. The polarization of all the phonon branches is presented.

The low-frequency modes $B_{2g}$ and $E_{2g}$ below 60 cm$^{-1}$ (denoted as $B_{2g}^2$ and $E_{2g}^2$) in $2H-MoS_2$, have no cousin modes in the monolayer. However, the two modes evolve into a series of shear (S) and compression (C) modes in FL $MoS_2$ spreading around 30 cm$^{-1}$, respectively [34-37]. These low-energy optical phonons are easily thermally activated at room temperature and believed to greatly affect the carriers mobility and thermal conductivity via electron-phonon scattering and phonon-phonon scattering as similar in FL graphene [38,39]. These interlayer S and C modes in FL $MoS_2$ show a strong anharmonic character and layer-dependent frequency shift, and have been observed experimentally only recently [37]. We also predict a large Grüneissen parameter of the low frequency acoustic modes in monolayer as shown below.

The phonon dispersion and density of states (DOS) of the monolayer $MoS_2$ are shown in Figure 1b. There are three acoustic branches: transverse acoustic (TA), longitudinal acoustic (LA), and out-of-plane transverse acoustic (ZA) branches, which are separated by a gap of around 50 cm$^{-1}$ below the non-polar transverse optical (TO) and longitudinal optical (LO) modes, labeled as $TO_1$ and $LO_1$, respectively. The $TO_2$, $LO_2$, and $ZO_1$ modes are three polar branches. The homopolar $ZO_2$ branch shows a nondispersive behavior, accompanying with a breathing mode eigenvector (see $A_1'$ mode in Figure 1a). In polar semiconductor or insulators, each IR active mode (polar mode) displays the LO/TO splitting due to the coupling of the lattice to the polarization field created by the polar mode in the long-wavelength limit. For bulk $2H-MoS_2$, the Born effective charges (BEC) of Mo and S are small [33] and the polarized fields associated with the two IR modes ($A_{2u}$ and $E_{1u}$) are weak (only leading to a 2 cm$^{-1}$ LO/TO splitting [28]) due to a small mode oscillator strength [40,41]. For monolayer $MoS_2$, the electronic screening is weaker than 3D case and the splitting will be even smaller and thus neglected here.

Table 1. Phonon parameters of single-layer MoS$_2$ at Γ for three acoustic modes (labeled according to polarization) and four optical modes (labeled according to symmetry assignment of irreducible representation of D$_{3h}$ point group): frequency (ω), Grüneissen parameter (γ) determined by biaxial strain, and relaxation time (τ). Note that γ and τ for the three acoustic modes (ZA, TA, and LA) are shown at the cutoff frequency of 0.05 TH$_Z$.

| | ZA | TA | LA | E″ | E′ | A$_1'$ | A$_2''$ |
|---|---|---|---|---|---|---|---|
| ω(cm$^{-1}$) | 0 | 0 | 0 | 298.2 | 402.7 | 423.9 | 490.5 |
| | | | | 289.2[a] | 391.7[a] | 410.3[a] | 476.0[a] |
| γ | 159.71 | 58.63 | 25.31 | 0.42 | 0.54 | 0.20 | 0.44 |
| | | | | 0.54[b] | 1.06[c], 0.21[b], 0.6[d] | 0.21[b] | 0.53[b] |
| τ(ps) | 6.93 | 22.17 | 48.21 | 16.84 | 5.50 | 38.36 | 5.72 |
| | | | | | 5.1[e] | | 2.1[e] |

[a]Ref. [32];

[b]Ref. [46];

[c,d]γ obtained from Ref. [47] and Ref. [10], respectively, by measuring the shift of the Raman peak through applying uniaxial strain;

[e]Relaxation time (τ) measured for the bulk 2H-MoS$_2$ phase of the two IR-active modes: E$_{1u}$ and A$_{2u}$ modes corresponding respectively to the E′ and A$_2''$ modes of the monolayer case, by analyzing the damping constant of the image dielectric spectrum in Ref. [56].

In Figure 2, we calculate the phonon group velocity $V_n = d\omega_n/dq$ for the $n$th branch along the Γ-M high symmetrical line. The $V-q$ and $V-\omega$ relationships are plotted in the Figure 2a and b, respectively. For the TA and LA modes along the Γ-M direction, the sound velocities at long wavelength limit are about 693.5 and 1108.8 m/s, respectively, much smaller than the graphene case of 3743 (TA) and 5953 (LA) m/s [23]. For LA and TA modes, the group velocities drop dramatically with increasing phonon frequency, while for ZA mode, its group velocity increases with the frequency, reaches maximum at the q point sitting at the middle of the Γ-M line, then decreases and finally reaches to zero at the zone edge.

### B. Grüneissen parameter

The Grüneissen parameter (γ), which provides information on the anharmonic interactions between the lattice waves and the degree of the phonon scattering, is calculated by dilating the lattice with ±0.5% of biaxial strains. For 2D materials, the γ of each phonon mode at $q$ point with $s$ polarization is given by [42,43]:

$$\gamma_{\mathbf{qs}} = -\frac{a}{2\omega_s(\mathbf{q})}\frac{d\omega_s(\mathbf{q})}{da} \qquad (3)$$

where $a$ is the relaxed equilibrium lattice constant of 3.09 Å. In Figure 3, the values of γ for the acoustic and optical modes of MoS$_2$ are plotted to reflect the mode-dependent strength of anharmonicity.

The values of γ for all the acoustic and optical modes at Γ point are compiled in Table 1. The values of γ for the optical modes are 0.42 (E″), 0.54 (E′), 0.20 (A$_1'$), and 0.44 (A$_2''$), respectively, consistent with the measured value of Raman active E′ of 0.6 in Ref. [10]. It is interesting to find that all the modes show a positive value of γ, indicative of a normal behavior of softening frequencies with expanding the lattice host [44]. The positive γ indicates a positive coefficient of thermal expansion of monolayer MoS$_2$ even at low temperatures where only the acoustic modes are excited. This is in contrast to the negative thermal expansion observed in graphene at low temperatures due to the negative γ of the ZA mode [45]. The issue of thermal expansion is critical for two-dimensional devices as the different thermal expansion coefficients between samples and substrate can result in strain, which may affect the performance and reliability for electronic device applications. Our results of Grüneissen parameter γ show that monolayer MoS$_2$ is a credible alternative to graphene in some applications when a positive thermal expansion monolayer material is preferred. The underlying reason for the difference in thermal-mechanical relationship between MoS$_2$ and graphene may be due to the suppression of the bending mode in MoS$_2$ owning to its sandwiched structure. While the γ of optical modes were reported previously [10,46,47], the γ for the acoustic modes is still unknown. The values of γ for the three acoustic modes at Γ point are 159.7, 58.6 and 25.3 for ZA, TA, and LA, respectively, which are overwhelmingly larger than those of the optical modes within the whole Brillouin zone. These results show that the anharmonicity is weaker for optical modes than acoustic modes.

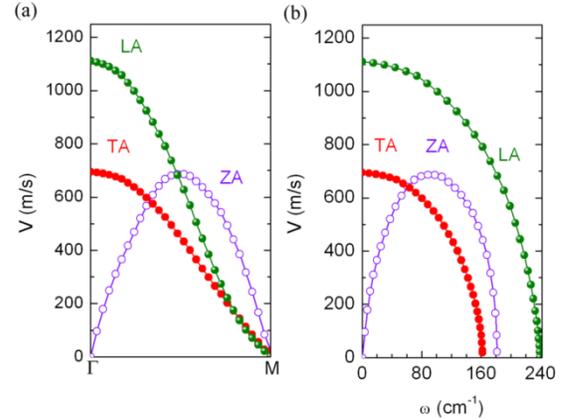

FIG. 2. Wavenumber (a) and frequency (b) dependent phonon group velocity of acoustic phonon modes along Γ-M direction.

The variation of the frequencies for modes at Γ point with compressive and tensile strains is plotted in Figure 3c, where the frequency shift with strain (δ) is defined as $\Delta\omega(\delta) = \omega(\delta) - \omega(0)$. The different slopes of the δ-Δω curves reflect the different stiffening or softening behavior of each phonon mode under strain. For A$_1'$ mode, the slope is the smallest among all the modes, consistent with the smallest γ of ZO$_2$ branch within the whole Brillouin zone (Figure 3b). Eigenvector of this mode shows that the S atoms vibrate in counter phase in direction normal to the plane (Figure 1) and the Mo plane keeps stationary. Thus the frequency is relatively insensitive to the in-plane strain, but quite sensitive to the disturbance normal to the plane like increasing layers [32], electronic doping or chemical doping above the planes. In contrast, the E′ mode involves the in-plane vibration, and thus is more sensitive to the in-plane strain. Note here that the trends for these two modes are reversed in the case of doping on the layer, where the $A_1'$ mode shows significant softening behavior whereas the $E'$ mode remains nearly constant [12].

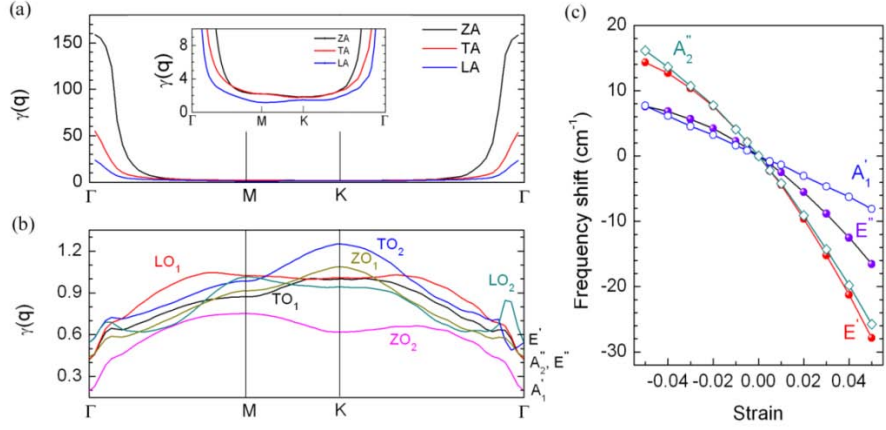

FIG. 3. The Grüneissen parameter of acoustic (a) and optical (b) modes of monolayer $MoS_2$, the inset in (a) is a close-up view. (c) Frequency shift for optical modes under biaxial strain in the sheet.

Our study of the larger slope of $E'$ mode ($E_{2g}$ for bulk $MoS_2$) than the $A_1'$ mode ($A_{1g}$ for bulk $MoS_2$) with strain is consistent with previous measurements [46,47]. Assuming a linear expansion of the sample with increasing temperature, the frequency shift can also be detected from the evolution of the Raman spectrum for samples at different temperatures. The positive $\gamma$ is consistent with the redshift of the Raman-active modes with temperature observed in experiments [9,19,48-50].

### C. Relaxation time and mean free path based on Umklapp process

The phonon relaxation time in real materials can be obtained by the combination of various scattering processes, such as phonon-phonon Umklapp scattering, boundary scattering and defects scattering. By improving sample quality, extrinsic scatters such as defects and grain boundary can be removed. In the present work, we focus on intrinsic phonon relaxation process in monolayer $MoS_2$ sheet, thus boundary scattering is not considered. The intrinsic phonon relaxation time associated with phonon-phonon Umklapp scattering was derived using an expression given by Klemens based on the time-dependent perturbation theory [51,52], but introducing separate lifetimes ($\tau_{\mathbf{q}s}$) for different phonons branches:

$$\frac{1}{\tau_{\mathbf{q}s}} = 2\gamma_{\mathbf{q}s}^2 \frac{k_B T}{M v^2} \frac{\omega_{\mathbf{q}s}^2}{\omega_m} \quad (4)$$

where $M$ is the atomic mass and $\omega_m$ is the Debye frequency, $T$ is the temperature, $k_B$ is the Boltzmann constant, and $v$ is the averaged sound velocity given by the relation $2/v^2 = 1/v_{LA}^2 + 1/v_{TA}^2$ around the zone-center. It should be noted that a cutoff frequency ($\omega_c$) for the long-wavelength acoustic phonons should be adopted in this formula to avoid the divergence issue [53,54]. In the previous study on bulk graphite, Klemens et al. [53] chose the frequency of the $ZO'$ mode at $\Gamma$ as the $\omega_c$ (around 4 THz) based on the assumption of the onset of cross-plane coupling and heat transport at this frequency. However, for monolayer $MoS_2$, this assumption cannot be applied as there is no $ZO'$ mode in the monolayer case. By non-dimensionalizing the data based on Umklapp scattering rates of phonons and assuming a constant sound velocity within the whole Brillouin zone, Freedman et al. recently showed an universal MFP spectrum in crystalline semiconductors and revealed that approximately 90% of thermal con-ductivity is contributed from phonons with MFPs in the range between 1 and 200 times of the MFP of acoustic phonons at the Brillouin zone edge [55], covering the phonon frequencies between 0.5% and 100% of the frequencies of acoustic phonons at the Brillouin zone edge. Based on this scenario, $\omega_c$ of 1.2 cm$^{-1}$ is selected here which is 0.5% of the frequency of LA phonon at M point (~240 cm$^{-1}$).

Figure 4a gives the frequency dependence of relaxation time of LA, TA and ZA phonons in monolayer $MoS_2$ sheet. For these acoustic phonon branches, the relaxation time decreases sharply as the frequency of phonons increases. The calculated $\tau(\omega_c)$ for both acoustic and optical branches are compiled in Table 1. The values for the ZA, TA, and LA modes are found to be 6.93, 22.17, and 48.21 ps, respectively. The ZA mode shows the shortest relaxation time as an indication of the strongest phonon scattering, being consistent with the largest Grüneissen parameter.

For the four optical modes, the relaxation times are 16.84 ps ($E''$), 5.50 ps ($E'$), 38.36 ps ($A_1'$), and 5.72 ps ($A_2''$). By fitting the classical dielectric oscillators with the measured IR reflectivity, Sun et al. recently obtained the relaxation times for the IR modes $E'$(5.1 ps) and $A_2''$ (2.1 ps) [56]. With consideration of the phonon-phonon Umklapp scattering, our predicted $\tau$ of the in-plane $E'$ mode is in good agreement with the measured value (5.1 ps), whereas for the $A_2''$ mode the $\tau$ is 2-fold larger than the experimental value. For real materials, the phonon relaxation time can be decomposed into contributions from electron-phonon, phonon-substrate and phonon-phonon Umklapp scattering. For the out-of-plane vibrating $A_2''$ mode, the contribution from electron-phonon and phonon-substrate coupling is more significant than that in the in-plane $E'$ mode due to the $d_{z^2}$ character in the frontier orbitals [12]. This may account for the calculated larger phonon relaxation time for the $A_2''$ mode with only considering the Umklapp scattering.

For the Raman active phonons, unfortunately, there is no measured value of the lifetime for comparison. Our result shows that the nonpolar optical $E''$ (homopolar $A_1'$) mode possesses a much longer $\tau$ (three and seven times, respectively) than those of the IR-active $E'$ and $A_2''$ polar modes. This indicates that the thermal scattering in the optical modes mainly occurs through the polar modes.

A fundamental understanding of the phonon MFP is essential to identify the phonon $\kappa$ of two-dimensional materials. Based on the mode-dependent $\tau$ and group velocity, we obtain the Umklapp scattering limited MFP of monolayer $MoS_2$, which is a key quantity for understanding the size-scaling characteristics of $\kappa$. The phonon MFP for mode at $q$ point with $s$ polarization is defined as $\lambda_{\mathbf{qs}} = V_{\mathbf{qs}}\tau_{\mathbf{qs}}$. It is well known that the acoustic modes contribute to most of the $\kappa$ due to their relatively larger velocities than those of the optical modes [23]. Therefore, we only analyze the acoustic modes phonon MFP. The frequency- and mode-dependent phonon MFP dominated by the Umklapp process is shown in Figure 4b. Similar to relaxation time, for all the phonon branches, MFP decreases obviously with increasing frequency, then increases and decreases slightly. Over the entire frequency regime, the MFP of LA mode is larger than those of TA and ZA modes, due to the combined effects from the frequency-dependent relaxation time and group velocities. To define an effective or frequency independent MFP for each branch, we employed the MFP at truncated frequency ($\omega_c$) as the dominated MFP: Beyond which all phonons experience non-ballistic transport. The largest dominated MFP among the zone-center acoustic modes is found to be the LA mode with a value of around 18.1 nm, which is much larger than that of the TA mode (5.0 nm). This finding is consistent with the value of 5.2 nm by the recent MD calculation performed by Liu et al. [15] Our calculation shows that the MFP of monolayer MoS$_2$ sheet is much smaller than that of the graphene (~775 nm) [20]. Since the sizes of most reported MoS$_2$ flakes are around 1μm [57-60], the thermal conduction in these samples is highly likely to be in the diffusive regime according to the predicted dominated MFP. Our study suggests that there appears to be a significant reduction of thermal conduction when the characteristic length, such as the size of sample or the size of grain, is reduced down to tens of nanometers as the grain boundary or edge boundary scattering becomes greatly enhanced.

### D. Thermal conductivity calculated from NEGF

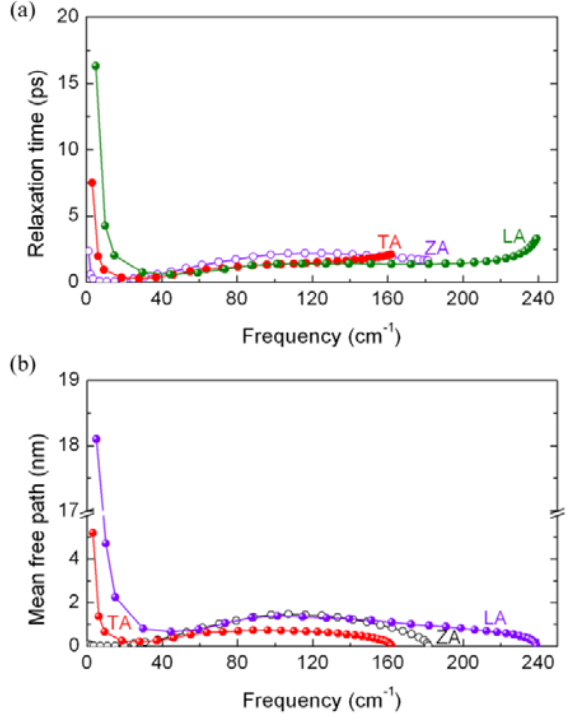

FIG. 4. (a) Frequency dependent relaxation time and (b) mean free path for the three acoustic modes (ZA, TA and LA) of monolayer MoS$_2$.

In this part, we calculate the thermal conductance and thermal conductivity of monolayer MoS$_2$ by using NEGF approach. The MoS$_2$ sheet is modeled by creating an orthogonal supercell containing 144 atoms with assigning the armchair direction as the thermal conducting direction. Periodic boundary condition is applied in both the heat current direction and in-plane transverse direction. As it is computationally highly demanding by DFPT calculation for this system, here we use the Vienna *ab initio* simulation package (VASP) package [61] with the Perdew-Burke-Ernzerhof functional (PBE) to obtain the Hessian matrix by shifting individual atoms in the supercell and determine the forces induced on all other atoms. Figure 5 shows the obtained phonon transmission spectrum. The difference in mode frequency between Figure 1 and Figure 5 due to the different approaches is small. The calculated thermal conductance is 1.06 nW/K for sample with width of 1.27 nm. Assumed the thickness of the MoS$_2$ sheet of 0.65 nm, the scaled thermal conductance, defined as thermal conductance per unit area $\sigma/S$, amounts to 1.28 nWK$^{-1}$nm$^{-2}$.

The thermal conductivity $\kappa$ of a finite sample is related to the thermal conductance $\sigma$ by $\kappa = \sigma l/S$, where $l$ is the length of the sample, and $S$ is its cross section. Within the ballistic regime ($l < \lambda$, $\lambda$ is phonon MFP), the $\kappa$ is linearly correlated with the length since the conductance $\sigma$ is length-independent [21]. In contrast, when the sample length is beyond mean free path ($l > \lambda$), the transport is diffusive and $\kappa$ is less sensitive to the variation of the length. It was proposed [21] that from the NEGF calculated ballistic thermal conductance, one may estimate the size-independent thermal conductivity of a sample with the relation $\kappa = \sigma\lambda/S$, where $\lambda$ is the phonon mean free path. Using this approach, the room temperature thermal conductivity of single-layer graphene calculated from NEGF is 3410 Wm$^{-1}$K$^{-1}$ [21], which is comparable with the experimental measurements [62]. Since thermal conductivity in semiconductors results mainly from phonons with long MFP [55,63], to derive the $\kappa$ of MoS$_2$, here we use the dominated MFP of zone-center LA mode ($\lambda$ = 18.1 nm) obtained in the lattice vibrational modes section. The room temperature (300 K) thermal conductivity of monolayer MoS$_2$ is found to be 23.2 Wm$^{-1}$K$^{-1}$. It is worth mentioning that if the phonon MFP of 5.2 nm (by the MD calculation in Ref. [15]) is adopted, the $\kappa$ is about 6.66 Wm$^{-1}$K$^{-1}$, which is in the same order of magnitude with the MD results of 1.35 Wm$^{-1}$K$^{-1}$ [15].

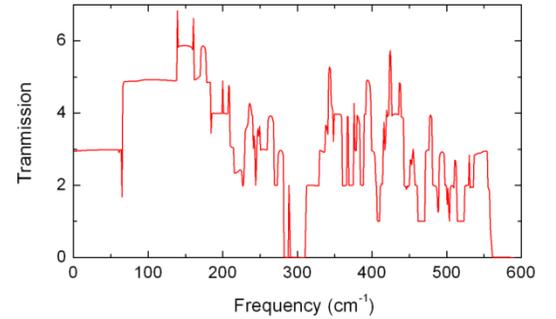

FIG. 5. Phonon transmission for monolayer MoS$_2$ calculated by the NEGF method.

Our predicted $\kappa$ of 23.2 Wm$^{-1}$K$^{-1}$ of MoS$_2$ is much smaller than the ultra-high thermal conductivity (4800-5600 Wm$^{-1}$K$^{-1}$) of graphene [62]. Since the scaled thermal conductance $\sigma/S$ value of 1.28 nWK$^{-1}$nm$^{-2}$ for MoS$_2$ is only slightly smaller than that of graphene (~4.1 nWK$^{-1}$nm$^{-2}$ from Ref. [64]) with the same width, the 100-fold reduction of $\kappa$ in MoS$_2$ arises from the much shorter MFP in MoS$_2$ compared with graphene.

Experimentally measured values of $\kappa$ of FL MoS$_2$ are from 0.4-1.59 Wm$^{-1}$K$^{-1}$ [17,18], to around 52 Wm$^{-1}$K$^{-1}$ by Raman spectroscopy approach [19]. This large discrepancy may originate from different sample quality, measurement methods and accuracy. Thus numerical simulations will definitely facilitate the understanding of heat conduction in monolayer MoS$_2$ sheet. MD has the obvious advantage that it does not rely on any thermodynamic-limit assumption, and is thus applicable to model nanoscale systems with real geometries. However, the calculated $\kappa$ may rely on the inter-atom empirical potentials used. The combination of first-principles calculation and NEGF method has been commonly used to describe the ballistic phonon transport. In the present work, by adopting the Umklapp phonon-phonon scattering limited MFP, our calculation provides an estimation of the intrinsic thermal conductivity. However, the intrinsic relaxation time of phonon-phonon Umklapp scattering is calculated from an empirical equation based on the time-dependent perturbation theory [51, 52]. Thus phonon scattering rates obtained fully from *ab initio* density functional theory calculations [65] without any adjustable parameters is still indispensable to determine the dominated phonon MFP.

## IV. CONCLUSION

In-depth understanding on the lattice vibrational modes and intrinsic thermal conductivity of $MoS_2$ is highly important to speed up its potential applications in nanoelectronics and thermoelectric energy devices. Based on the DFPT calculations, we have obtained the Grüneissen parameters and the Umklapp scattering limited relaxation time of phonons in monolayer $MoS_2$. The calculated phonon lifetime of the IR active modes is in good agreement with experimental results. The dominated phonon mean free path of the monolayer $MoS_2$ sheet is around 18.1 nm, about 30-fold smaller than that of graphene. By using the NEGF approach combined with the obtained mean free path, the intrinsic thermal conductivity of monolayer $MoS_2$ at room temperature is found to be around 23.2 $Wm^{-1}K^{-1}$, around two orders smaller than that of graphene, due to the significantly shorter phonon mean free path compared with graphene.

## ACKNOWLEDGMENTS

The authors gratefully acknowledge the financial support from the Agency for Science, Technology and Research (A*STAR), Singapore and the use of computing resources at the A*STAR Computational Resource Centre, Singapore.